\documentclass{svproc}

\usepackage{cite}
\usepackage{amsmath}
\usepackage{algorithmicx,xpatch}
\usepackage{algorithm}
\usepackage{algpseudocode}
\usepackage{graphicx}
\usepackage{textcomp}
\usepackage{xcolor}

\usepackage[utf8]{inputenc}
\usepackage[T1]{fontenc}
\usepackage{stmaryrd}
\usepackage{listings}
\usepackage{tikz}
\usepackage{setspace}
\usepackage{multirow}
\usepackage{url}
\usepackage{caption}
\usepackage{moresize}
\usepackage{multirow}

\usepackage{enumitem}
\usepackage{pifont}
\usepackage{booktabs}
\usepackage{etoolbox}
\usepackage{verbatim}
\usepackage{float}
\usepackage{ifthen}
\usepackage{scalefnt}
\usepackage{blkarray}
\usepackage{subfig}
\usepackage[mathscr]{eucal}

\def\mathbi#1{\textbf{\em #1}}

\usepackage{accents}

\makeatletter
\newcommand{\douwidehat}[2]{%
  \sbox0{$\m@th#1\widehat{\hphantom{#2}}$}%
  \sbox2{$\m@th#1x$}
  \sbox4{$\m@th#1#2$}
  \dimen0=\ht0
  \advance\dimen0 -.8\ht2
  \dimen2=\dp4
  \rlap{%
    \raisebox{\dimexpr\dimen0-\dimen2}{%
      \scalebox{1}[-1]{\box0}%
    }%
  }%
  {#2}%
}
\makeatother

\makeatletter
    \newcommand{\vast}{\bBigg@{3}}
    \newcommand{\Vast}{\bBigg@{3.5}}
    \newcommand{\vastt}{\bBigg@{4}}
    \newcommand{\Vastt}{\bBigg@{4.5}}
\makeatother

\def\BibTeX{{\rm B\kern-.05em{\sc i\kern-.025em b}\kern-.08em
    T\kern-.1667em\lower.7ex\hbox{E}\kern-.125emX}}

\title{RISC-V word-size modular instructions for Residue Number Systems}

\date{june 2024}

\author{
  Laurent-St\'ephane Didier \and Jean-Marc Robert}

\institute{
  {IMATH, Universit\'e de Toulon,
  Toulon, France \\
 laurent-stephane.didier@univ-tln.fr,
 jean-marc.robert@univ-tln.fr}
 }

\authorrunning{Laurent-St\'ephane Didier and Jean-Marc Robert}

\begin{document}

\maketitle

\begin{abstract}
Residue Number Systems (RNS) are parallel number systems that allow the computation on large numbers. They are used in high performance digital signal processing devices and cryptographic applications. However, the rigidity of instruction set architectures of the market-dominant  microprocessors limits the use of such number systems in software applications.

This article presents the impact of word-size modular arithmetic specific RISC-V instructions on the software implementation of Residue Number Systems. We evaluate this impact on several RNS modular multiplication sequential algorithms. We observe that the fastest implementation uses the Kawamura \textsl{et. al.} base extension.
Simulations of architectures with GEM5 simulator show that RNS modular multiplication with Kawamura's base extension is 2.76 times faster using specific word-size modular arithmetic instructions than pseudo-Mersenne moduli for In Order processors. It is more than 3 times for Out of Order processors.
Compared to x86 architectures, RISC-V simulations show that using specific instructions requires 4.5 times less cycles in  In Order processors and 8 less in Out of Order ones.
\end{abstract}


\section{Introduction}

This paper deals with modular multiplication of large numbers, and its applications. This operation is widely used in cryptographic computations such as RSA encryption/decryption, Elliptic Curve Cryptography, and more recently, the SIKE post-quantum protocol. Implementations of this operation aim to provide fast and secure computation. In the context of software implementation, the state-of-the-art cryptographic libraries such as OpenSSL make use of multi-precision approaches. In this work, we explore the use of the non-positional number system Residue Number Systems (RNS) in this context. We target the RISC-V platform and software implementations of this approach in order to evaluate the efficiency and the potential improvement offered by extension of the instruction set, in particular specific word-size modular arithmetic instructions. This research is based upon simulation of RISC-V platforms, and is prelude to future hardware implementations and experiments on real platforms.

Residue Number Systems (RNS) are non-weighted carry-free number systems which arithmetic is done over parallel finite rings. Such systems can be used in high performance signal processing \cite{cmzt2015}, fault-tolerant computing \cite{tay2017fault}, convolutional neural networks \cite{valueva2020application} and
cryptographic applications such as RSA cryptosystems \cite{BajardI04},  homomorphic cryptography \cite{behz_16}, elliptic curves\cite{abs_12}. These systems also have interesting leak-resistant properties \cite{bde_13} \cite{lesavourey2017efficient}.
Because of their parallel property, they are suitable for vector implementations. However, the software implementation of RNS applications suffers from the cost of the word-size modular arithmetic \cite{didier2022software}.

RISC-V is an open standard Instruction Set Architecture (ISA) that aim to provide a free and open ISA suitable for processor designs. It was pioneered at University of California at Berkeley in 2010. Now, this initiative is supported by the RISC-V association that regroups many industrials and academics interested in such a collaboration.
The RISC-V standards organization continuously introduces new ISA extensions to meet the needs of advanced computing. 
The ISA itself is designed according to the reduced instruction set computing (RISC) principles. It consists of a small base integer instruction set with several sets of modular extensions (integer multiplication, floating point operations, etc.). This ISA also has  a dedicated space for future or custom extensions. There is a lot of novel research for customized extensions of RISC-V for specific scenarios \cite{cui2023risc}. 

This offers the opportunity to introduce new arithmetic-related  instructions. In the context of post-quantum cryptography, some extensions for lattice-based crypto-protocols have been proposed to improve the NTT computation  \cite{fritzmann2020risq} \cite{karabulut2020rantt}, LAC scheme \cite{fritzmann2020extending} or for computations in finite fields \cite{alkim2020isa}. The extension of ISA can also be useful for more \textsl{exotic} number systems.
Some extensions have been proposed for POSIT arithmetic \cite{tiwari2021peri} \cite{mallasen2022percival} and bit-slicing computations \cite{kiaei2023architecture}.

In this article we show the impact of word-size modular arithmetic instructions on the RNS operations. We target the modular multiplication which is a frequent and expensive operation in several cryptographic \cite{menezes2018handbook}. In RNS, this operation requires the conversion between two bases that can be computed through several ways.

\paragraph{Related works}

As mentioned above, RNS are widely used in various contexts in hardware implementations and are based on word-size modular operations \cite{mohan2017rns}. This is motivated by the inherent property of these systems to be parallelizable. In our case, since we are targeting software implementations, there are very few works in this topic. Some improvements to the software implementation of word-size modular operations have been proposed to use actual processor arithmetic units \cite{Plantard21} \cite{aoki2022efficient}. Some of these improvements are also useful in lattice-based cryptosystems using Number Theoretic Transform (NTT) \cite{huang2022improved} \cite{huang2023yet}. However the word-size modular operations remain an bottleneck in software implementations.

A comparative study that highlights this phenomenon has been proposed by Didier \emph{et al.} in\cite{didier2022software}. Their work explores the software RNS implementation of modular multiplication for various modulus precision, from 400 to 3251 bits, on \texttt{x86-64} platform. This work compares sequential (using the classical instruction set) and parallel (\texttt{AVX512}) implementations with the multi-precision \texttt{GMP} library \cite{gnu_mp}. This work concludes on the costly impact of the word-size modular operation and its penalty on the performance. Our work presented here attempts to address this issue on the software sequential RNS implementations, on RISC-V platforms. 

\paragraph{Contributions}
We evaluate the benefit of the use of processors having specific instructions for the word-size modular operation and compare it with regular software implementation. We target RISC-V ISA. Our evaluations depend on several parameters:
\begin{itemize}
    \item the word-size modular operation method for the elementary RNS operations,
    \item the conversion methods between two RNS bases,
    \item the size of the RNS base,
    \item the RISC-V processor configuration,
    \item the cost of RISC-V specific word-size modular operation instructions.
\end{itemize}
We ran nearly 3,000 simulations with the GEM5 simulator which is a modular platform for computer-system architecture research \cite{binkert2011gem5}.

\paragraph{Paper organisation}
The background on RNS and the RNS modular multiplication are reminded in section \ref{sec:RNS}.  Our new instructions are described in section \ref{sec:instruction}, the experiment parameters are described in section \ref{sec:expermiment}. We show the experimental results in section \ref{sec:results}.

\section{Residue Number System}\label{sec:RNS}

Residue Number Systems are non positional integer number systems that are based on the Chinese Remainder Theorem ~\cite{gar_56, taylor84, knuth2014art }.
In such a system, an integer $x$ is represented by its remainders $x_i =x \bmod m_i$. The values $m_i$ are relatively prime numbers. The set $\mathcal{B}_m=\{m_{1}, m_{2}, \ldots, m_{n}\}$ forms the RNS base composed of $n$ channels. The moduli $m_i$ are usually chosen with the width $w$ that corresponds to the target architecture word-size. We denote $\mathbi{M}$ their product. The advantage of such a number system is that additions, subtractions and multiplications can be performed in parallel on each channel: 

\[
	z_i = x_i \odot y_i \bmod m_i \mbox{ where } \odot \in\{+, -, \times\}
\]

\paragraph{Conversions} The forward conversion to RNS is simply a modular operation on each base channel. The backward conversion can be done through different ways. The Chinese Remainder Theorem provides a computation formula in the target number system~\cite{knuth2014art}:

{\small\begin{equation}\label{CRT}
	x = \left| \sum_{i=1}^{n} x_i \left(\frac{\mathbi{M}}{m_i}\right)_{m_i}^{-1} \mathbi{M}_i \right|_{\mathbi{M}} =  \sum_{i=1}^{n} x_i \left(\frac{\mathbi{M}}{m_i}\right)_{m_i}^{-1} \mathbi{M}_i -k\cdot \mathbi{M}
\end{equation}}
where
{\small\[ 
\mathbi{M}_i\times \left(\frac{\mathbi{M}}{m_i}\right)_{m_i}^{-1} \equiv 1 \pmod {\mathbi{M}}
\]}

The main drawback of this approach is that the values used in this sum are large. 

An other method consists of the conversion into the Mixed Radix System. This requires modular computations on $w$-bit integers only. In this positional system, an integer $x_{MRS}$ is as follows:
\[x_{MRS}=x'_{0}+x'_{1}m_{0}+x'_{2}m_{0}m_{1}+\cdots+
x'_{n-1}\prod_{i=0}^{n-2}m_{i}\]
This conversion requires $\mathcal O(n^2)$ operations on $w$-bit operands and needs $\mathcal O(n^2)$ constants~\cite{szabo-tanaka67}.

A trade-off between these two methods has been proposed by Kawamura \textsl{et al.}~\cite{Cox-Rower00}. It is based on equation (\ref{CRT}) and consists of the estimation of $k$ with approximate values through $\mathcal O(n)$ operations on small values with $\mathcal O(n)$ constants. The approximate values are $w$-bit integers.

\paragraph{Base extension}
The base extension is the conversion of an RNS number from one RNS base to another. This  consists of a backward conversion and a forward conversion to the targeted RNS base. Both operations are interleaved in order to minimize storage of intermediate values. 

The first base extension has been proposed by Szabo and Tanaka. It is based on the mixed-radix conversion~\cite{szabo-tanaka67}. Shenoy and Kumaresan suggested to compute the value $k$ in equation (\ref{CRT}) using an extra modulus $m_e$~\cite{shenoy-kumaresan89}. 
If $k$ is known, then it is possible to compute equation (\ref{CRT}) in the target RNS base. Similarly, Kawamura \emph{et al.}~\cite{Cox-Rower00} proposed a conversion method based on their  approximation of $k$.

\paragraph{RNS Modular multiplication}
 \label{par:RNSmodmult}
In RNS, the modular multiplication is derived from the Montgomery multiplication~\cite{montMult_85} and requires base extensions~\cite{posch1995modulo, bajard1998rns}. 
It is summarized in Algorithm~\ref{Alg:RNS_mod_mul}. 
In~\cite{BajardI04}, the authors remark that if the dynamic range of base $\mathcal{B}_{m}$ is large enough, then it is not necessary to completely compute the first base extension which can be approximated. For the second one, they use the Shenoy-Kumaresan method~\cite{shenoy-kumaresan89}. In the multiplication described in~\cite{Cox-Rower00} both extensions are Kawamura's.

In our implementations of Algorithm~\ref{Alg:RNS_mod_mul}, we chose $\mathcal{B}_m$ and $\mathcal{B}_{m'}$ in order to use the Bajard-Imbert~\cite{BajardI04} first extension at step 3. For the second extension at step 7, we use the Szabo-Tanaka method~\cite{szabo-tanaka67} or Kawamura \emph{et al.} method~\cite{Cox-Rower00}.

\begin{algorithm}[ht]
	\caption{RNS Modular Multiplication} \label{Alg:RNS_mod_mul}
	\begin{algorithmic}[1]
		\Require $x$ in $\mathcal{B}_m$ and $\mathcal{B}_{m'}$; 
		 $y$ in $\mathcal{B}_m$ and $\mathcal{B}_{m'}$ such that $x<2p$ and $y<2p$.
		 
		 \hspace{-1.3cm}\textbf{Precomputation:} $-p^{-1}$ in $\mathcal{B}_{m}$; $p$ in $\mathcal{B}_{m'}$; $\mathbi{M}^{-1}$ in  $\mathcal{B}_{m'}$
		\Ensure  $z=x\times y \times \mathbi{M}^{-1} \bmod p$ in $\mathcal{B}_m$ and $\mathcal{B}_{m'}$ such that  $z<2p$.		
        \State $s \leftarrow x\times y$ in  $\mathcal{B}_{m'}$ and $\mathcal{B}_{m}$
		\State $t \leftarrow s\times (-p^{-1})$ in  $\mathcal{B}_{m}$
		\State Base extension of $t$ from $\mathcal{B}_{m}$ to $\mathcal{B}_{m'}$
		\State $u \leftarrow t \times p$ in $\mathcal{B}_{m'}$
		\State $v \leftarrow s + u$ in $\mathcal{B}_{m'}$
		\State $w \leftarrow v \times \mathbi{M}^{-1}$ in $\mathcal{B}_{m'}$
		\State Base extension of $w$ from $\mathcal{B}_{m'}$ to $\mathcal{B}_{m}$
		\State \Return $w$
	\end{algorithmic}
\end{algorithm}

\section{New instructions for word-size modular arithmetic}\label{sec:instruction}

The RISC-V project aim to provide an open RISC instruction set architecture for processor design. This project was started in 2010 at the University of California at Berkeley. Compared to x86 and ARM ISA, the RISC-V Foundation allows some customizations in the ISA specification. Some opcodes are reserved for custom instructions.

In the RISC-V instruction set, the arithmetic operations are performed register to register. In the instruction format, the fields related to the register are always at the same place. The input registers are denoted \texttt{rs} and the output register is denoted \texttt{rd}. In this format, the seven least significant bits encode the instruction \textsl{opcode}. The standard provides for \textsl{custom} opcodes \cite{RiscvManual1} that we are using in our proposition.

\begin{figure}[ht]
    \centering
    \includegraphics[width=0.8\columnwidth]{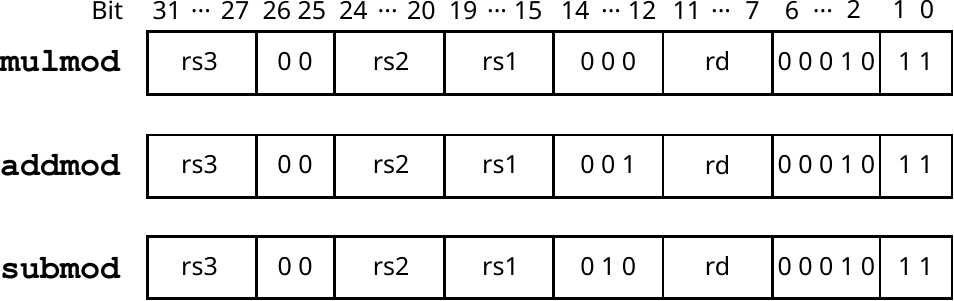}
    \caption{The instructions format for word-size modular arithmetic}
    \label{fig:instructions}
\end{figure}

We propose three new instructions for modular addition, subtraction and multiplication. The modular operations we implement require three inputs: the two operands and the modulus. Similarly to multiply-add vector instructions \cite{RiscvManualVector}, we use a third input register \texttt{rs3} which field is at bit 27-31 in the format instruction. 
The formats used for our word-size modular arithmetic instructions are summarized in Fig. \ref{fig:instructions}.

We provide hereafter the corresponding instruction operations:
\begin{enumerate}
    \item \texttt{mulmod rd, rs1, rs2, rs3}
\begin{algorithmic}
\State $rd \gets rs1 \times rs2 \bmod rs3$ 
\end{algorithmic}
    
    \item \texttt{addmod rd, rs1, rs2, rs3}
\begin{algorithmic}
\State $rd \gets rs1 + rs2 \bmod rs3$ 
\end{algorithmic}
    \item \texttt{submod rd, rs1, rs2, rs3}
\begin{algorithmic}
\State $rd \gets rs1 - rs2 \bmod rs3$ 
\end{algorithmic}

\end{enumerate}
\normalsize

These instructions are used through intrinsic C functions. As an example, Figure \ref{fig:intrinsic} shows the intrinsic function used for the modular addition. Our new instructions have been added to the  \texttt{gcc} cross-compiler from the RISC-V GNU Compiler Toolchain~\cite{riscvToolChain}. We also checked our compiled code with the Spike RISC-V ISA Simulator~\cite{riscvToolChain}. 

\begin{figure}[ht]
{\small
\begin{verbatim}{C}
inline static int64_t addmod(int64_t a, 
                  int64_t b, int64_t m)
{
  int64_t res;
  asm volatile
  (
    "addmod   %[z], %[w], %[x], %[y]\n\t"
    : [z] "=r" (res)
    : [w] "r" (a), [x] "r" (b), [y] "r" (m)
  );		
  return res;
}
\end{verbatim}
}
    \caption{Intrinsic C function for word-size modular addition}
    \label{fig:intrinsic}
\end{figure}

\section{Experiments}\label{sec:expermiment}

We used and adapted the RNS library from \cite{didier2022software} that we compiled with \texttt{gcc} with the \texttt{-O3} option. The binaries have been run within the GEM5 simulator and the number of cycles have been counted with the \texttt{rdcycle} for the RISC-V architecture and \texttt{rdtsc} for the x86 architectures. We used the evaluation protocol described in \cite{didier2022software}.

\subsection{Simulator}
The GEM5 simulator is an open source computer architecture cycle-level full-system simulator \cite{binkert2011gem5}. It is composed of a simulator core and parameterized models for a wide number of components such as in-order and out-of-order processors, DRAM or cache memories. 
It has been designed in order to effectively and efficiently emulate the behavior of modern processors. Amongst the various instruction set architectures, RISC-V \cite{roelke2017risc5} and x86 ISA are available \cite{binkert2011gem5}.

This simulator appears to be accurate enough in order to quickly evaluate ISA extension, architectural choices and discard or select solutions for further investigations \cite{butko2012accuracy, endo2014micro, rawat2017vector, walker2018hardware, akram2019validation, qiu2023gem5tune}. 
We used GEM5 version 22.1 for our experiments.   

\subsection{GEM5 simulator parameters}

The simulated RISC-V ISA uses 64-bit words. We chose a two levels 4-way associative cache architecture. The L1 cache is split in Data and Instruction parts of 32kb each. The L2 cache is set to 256kb. The simulated processors can access 2Gb dual channel DDR4\_2400 RAM. The clock frequency is set to 1Ghz. These parameters are the same for all our simulations.

The GEM5 simulator offers In Order (Minor) and Out of Order (O3) processors. Both models allow to tune some parameters such as the operators, their delay or, in case of Out of Order processors, their number. The Minor model is a four-stage pipeline In Order processor, while the O3 model is a five-stage Out of Order processor. The operators are pipelined, except for the divisor and modulo.

\subsection{Evaluated functions}
We used several algorithmic parameters in our evaluation. The number of RNS channels is between 8 and 64 in steps of 8, which corresponds to moduli from 512 up to 4096 bits for the RNS modular multiplication. The word-size modular operations have been implemented with three different modulo operator (mod.):

\begin{enumerate}
    \item The naive implementation provided by the C language, that is:

    \begin{verbatim}{C}
        int64_t res, a, m;
        res = a%m;
    \end{verbatim}

 In case of processors without \texttt{DIV} instruction, i.e. our situation, the code will be compiled as a sequence of instructions computing a full division, and this is very costly.

 In the sequel, this version is named \emph{Modulo} or \emph{mod}.

 \bigskip
 
    \item Pseudo-Mersenne moduli (PM) \cite{Plantard21}:

In this case, the moduli are of the form $m_i = 2^w - c_i$ where $w$ is the word size in bits and $c_i << 2^w$. Thus, one has $2^w \equiv c_i \bmod m_i$, and writing $a = a_l + 2^wa_h$ leads to $a\equiv a_l + c_ia_h \bmod m_i$. Though $c_ia_h$ may overflow $2^w$ bit, it is enough to repeat this process three times to get a fully reduced $\bmod~ m_i$ value of $a$ (see \cite{Plantard21}).

This corresponds to the Algorithm \ref{Alg:Mersenne} for the modular reduction.
\begin{algorithm}[ht]
	\caption{Pseudo-Mersenne modular reduction} \label{Alg:Mersenne}
	\begin{algorithmic}[1]
    \Require $a = a_l + a_h\times 2^w$ and $c_i$, precomputed $mask = 2^w-1$
    \Ensure $r\gets a\bmod m_i$
    \State $up\gets a_h$
    \State $lo\gets a_l$
    \State $t\gets c_i\times up$
    \State $up_2\gets t>>w$~// right $w$-bit shift 
    \State $lo_2\gets t\& mask$~// $w$-bit masking
    \State $t\gets t+lo+lo_2+c_i\times up_2$
    \State $up_3\gets t>>w$
    \State $lo_3\gets t\& mask$
    \State $t\gets lo_3+c_i\times up_3$
    \State \Return $r\gets t$
    \end{algorithmic}
\end{algorithm}

\medskip
This approach takes advantage of the special form of the moduli. It is much more efficient than the previous case, trading the computation of a full division by three multiplications by the small constant $c_i$ and a few additions, shifts and masking operations.

In the sequel, this version is named \emph{Pseudo-Mersenne} or \emph{PM}.

\medskip

\item Using our new instructions (Inst.):

In this case, the cost of the modular computation corresponds to the one of the single corresponding instruction.

In the sequel, this version is named \emph{Instruction} or \emph{Inst.}
    
\end{enumerate}

The evaluated RNS modular multiplication is described in Algorithm \ref{Alg:RNS_mod_mul}. In this algorithm, the most expensive function is the base extension. 
We have tested two variants of the RNS modular multiplication. While we use the Bajard-Imbert~\cite{BajardI04} first extension at step 3 for both variants, the second base extension at step 7 is either the Szabo-Tanaka in the first one~\cite{szabo-tanaka67} or the Kawamura \textsl{et al.} method in the second one~\cite{Cox-Rower00}.

In the sequel, the first variant is named \emph{Szabo-Tanaka} or \emph{ST} and the second variant is named \emph{Kawamura} or \emph{K}. These parameter abbreviations are used in the next figures of section \ref{sec:results}.

This leads to a total of six configurations:

\begin{itemize}
    \item Modulo Szabo-Tanaka
    \item Modulo Kawamura
    \item Pseudo-Mersenne Szabo-Tanaka
    \item Pseudo-Mersenne Kawamura
    \item Instruction Szabo-Tanaka
    \item Instruction Kawamura
\end{itemize}

\subsection{Experimentation parameters} To evaluate the impact of the use of specific modular operation instructions on RNS, we carried out simulations  varying some parameters. We evaluated the combination of the algorithmic parameters described in the previous paragraph.
The number of RNS channels ranges from 8 to 64 by 8 steps. The delay of the integer ALU is set to 1. As a consequence the additions are computed with a delay of 1. The delays of the integer multiplier unit varies from 3 to 4. They are set between 2 and 4 for the modular adder and between 4 and 9 for the modular multiplier, which is referred to as \textsl{long delays} case in the sequel. Finally, we simulated In Order (IO) and Out of Order (OoO) processor models.

\section{Results}\label{sec:results}

In this section, we first provide a global overview of the simulations. We afterward present the results of the simulations fetching the configurations mentioned in section \ref{sec:expermiment}.

 {\small
\begin{table*}[ht]
    \centering
    \begin{tabular}{|l|rl|rl|c|c|}
    \hline
   Processor & Modular op. & Base extension & Modular op. & Base extension & Ratio \\
    \hline
    \hline
    \multicolumn{6}{|c|}{\texttt{mulmod} delay: 4 \hspace{1cm} \texttt{addmod} delay: 2}\\
\hline
  IO & Modulo & Szabo-Tanaka & Instruction & Kawamura \textsl{et al.} &  7.49 \\
  IO &  Modulo & Szabo-Tanaka &  Instruction & Szabo-Tanaka &  5.46 \\
  IO &  Modulo & Szabo-Tanaka &  Pseudo-Mers. & Szabo-Tanaka & 2.08 \\

  IO & 	 Pseudo-Mers. & Szabo-Tanaka &  Instruction & Szabo-Tanaka & 2.63 \\
  IO & 	Pseudo-Mers. & Szabo-Tanaka &  Pseudo-Mers. & Kawamura \textsl{et al.} &  1.30 \\

  IO &  Pseudo-Mers. & Kawamura \textsl{et al.} &  Instruction & Kawamura \textsl{et al.} &  2.76 \\
  IO & Instruction & Szabo-Tanaka &  Instruction & Kawamura \textsl{et al.} & 1.37 \\

\hline
   OoO & Modulo & Szabo-Tanaka & Instruction & Kawamura \textsl{et al.} & 25.79 \\
  OoO & Modulo & Szabo-Tanaka &  Instruction & Szabo-Tanaka &  19 \\
  OoO & Modulo & Szabo-Tanaka &  Pseudo-Mers. & Szabo-Tanaka & 4.53 \\

	  OoO & Pseudo-Mers. & Szabo-Tanaka &  Instruction & Szabo-Tanaka & 4.19 \\
  OoO & Pseudo-Mers. & Szabo-Tanaka &  Pseudo-Mers. & Kawamura \textsl{et al.} &  1.86 \\

  OoO & Pseudo-Mers. & Kawamura \textsl{et al.} &  Instruction & Kawamura \textsl{et al.} &  3.06 \\
  OoO & Instruction & Szabo-Tanaka &  Instruction & Kawamura \textsl{et al.} & 1.37 \\
\hline
\hline
\multicolumn{6}{|c|}{\texttt{mulmod} delay: 9 \hspace{1cm} \texttt{addmod} delay: 4} \\
\hline
  IO & Pseudo-Mers. & Szabo-Tanaka &  Instruction  & Szabo-Tanaka &  1.87 \\

  IO & Pseudo-Mers. & Kawamura \textsl{et al.} &  Instruction & Kawamura \textsl{et al.} &  1.92 \\
\hline
  OoO & Pseudo-Mers. & Szabo-Tanaka &  Instruction  & Szabo-Tanaka &  2.74 \\

  OoO & Pseudo-Mers. & Kawamura \textsl{et al.} &  Instruction & Kawamura \textsl{et al.} &  2.30 \\
\hline
    \end{tabular}
    \caption{Speed ratio of 64-channel RNS modular multiplication with several word modular operations and base extensions, In Order (IO), Out of Order (OoO) RISC-V}
    \label{tab:perf_ratios}
\end{table*}
}

\subsection{Overview of the simulations}

Table \ref{tab:perf_ratios} shows the main and most significant results. The table is organised as follows:

\begin{itemize}
    \item The first rows consider the fastest delays used in our experiment for the word-size operations. The \texttt{mulmod} delay is 4,  the \texttt{addmod} delay is 2.
    \begin{itemize}
        \item In the In Order processor model, we first give the range of speed-ups for various configurations. For example, the ratio between the slowest version (Modulo, Szabo-Tanaka) and the fastest (Inst. Kawamura \textsl{et al.}) is 7.49.
        \item In the Out Of Order processor model, the performance hierarchy remains the same, however, the ratios are greater. The maximum speed-up is now 25.79, for the same versions as previously.
    \end{itemize}
    \item The last rows consider the longest delays for the word-size operations: \texttt{mulmod} delay: 9, \texttt{addmod} delay: 4. The ratios are lower, however, even in this case, the benefit of the specific word-size modulo instructions remains significant:
    \begin{itemize}
        \item The speed-up ratios of the Pseudo-Mersenne word-size modular reduction over the Instruction versions are nearly 2, in In Order processor model.
        \item The speed-up ratio between the Pseudo-Mersenne word-size modular reduction and the Instruction versions, in Out of Order processor model reaches 2.74 with the Szabo and Tanaka version.
    \end{itemize}
\end{itemize}

\begin{figure}[ht]
    \centering
\includegraphics[width=0.95\columnwidth]{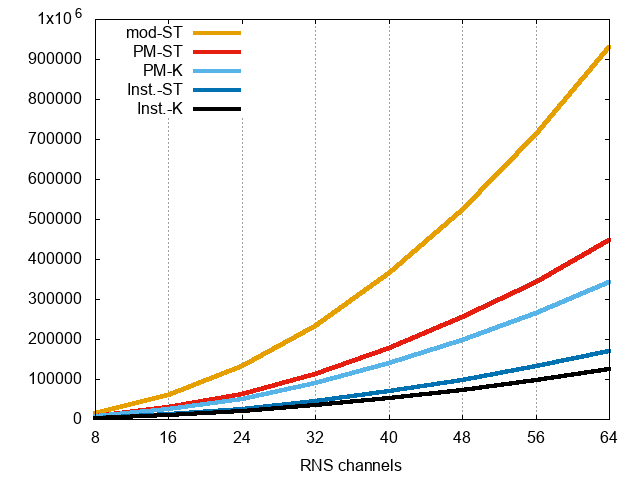}
    \caption{RNS modular multiplication timing in clock cycle number, with In Order RISC-V model, \texttt{mulmod} delay: 4, \texttt{addmod} delay: 2}
    \label{fig:Algo_MulMod}
\end{figure}

We do not provide all the timing values in clock cycle number. Nevertheless, Figure \ref{fig:Algo_MulMod} gives an overview of the orders of magnitude for the slowest architecture configuration, which is the RISC-V In Order processor model. The general quadratic behavior of Algorithm \ref{Alg:RNS_mod_mul} is observed for all configurations. In case of 64 RNS channels (4096-bit RNS modular multiplication) and for one single RNS modular multiplication, the slowest version (Modulo, Szabo-Tanaka) takes 928804 clock cycles and the fastest (Inst. Kawamura \textsl{et al.}) takes 149096 clock cycles (\texttt{mulmod} delay: 4, \texttt{addmod} delay: 2).

\subsection{RNS Modular multiplication algorithms}

We first compare the RNS modular multiplication algorithms which mainly depend on the base extension methods. We tested two variants for the second base extension at step 7 of Algorithm \ref{Alg:RNS_mod_mul}: the Szabo-Tanaka (ST)~\cite{szabo-tanaka67} and the Kawamura \textsl{et al.} (K)~\cite{Cox-Rower00} methods.  

The Figure \ref{fig:Algo_MulMod} summarizes the timing results expressed in clock cycles number for the In Order processor model. Without surprise, the measured delays regularly depend  on the number of RNS channels for all tested algorithmic parameter combinations.

\begin{figure}[ht]
    \centering
\includegraphics[width=0.95\columnwidth]{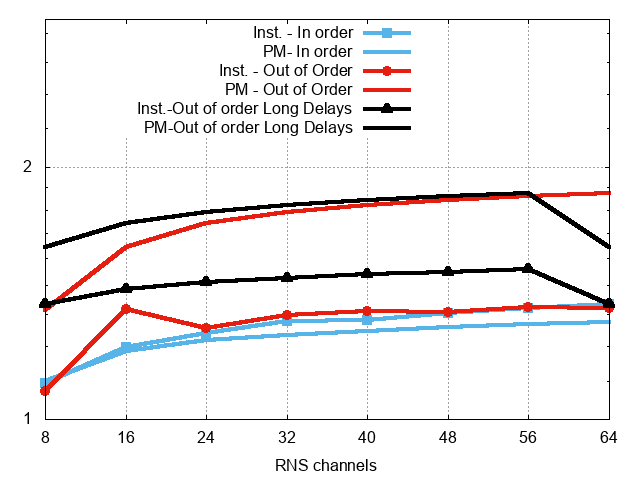}
    \caption{RNS modular multiplication Speed-Up, comparison of Kawamura \textsl{et al.}'s versus Szabo and Tanaka methods}
    \label{fig:SpeedUp_K-ST}
\end{figure}

Figure \ref{fig:SpeedUp_K-ST} highlights the comparison between the base extension methods for In Order and Out of Order processors with \texttt{mulmod} delay=4 and \texttt{addmod} delay=2 in blue and red lines respectively. The results of Out of Order processors with longer delays (\texttt{mulmod} delay=9 and \texttt{addmod} delay=4) are drawn in black. This figure shows the speed-up of the Kawamura's over Szabo-Tanaka method.
It reaches a maximum of 1.86 in the pseudo-Mersenne case, Out of order processor. The speed-ups seem to have a  weak correlation with the number of RNS channels.

The first part of the table \ref{tab:perf_synth} shows the speed-ups for 64-channel RNS focusing on the base extension comparison.
The fastest base extension is the Kawamura's method regardless of the method used for the word-size modular operations. Using the same modular operation method, the use of Kawamura's base extension is more than 30\% faster than Szabo-Tanaka's.

Finally, the best implementation that does not use our instructions is the combination of pseudo-Mersenne and Kawamura \textsl{et al.}'s method. We did not implemented the improvement of the word-size modular arithmetic provided in \cite{Plantard21} \cite{aoki2022efficient}, but the benefit seems to be close to the use of pseudo-Mersenne moduli.

\subsection{Word-size operations}

We now compare the performances of the RNS modular multiplication with respect to the word-size operation, in case of In Order processor model.

\begin{figure}[ht]
    \centering
\includegraphics[width=0.95\columnwidth]{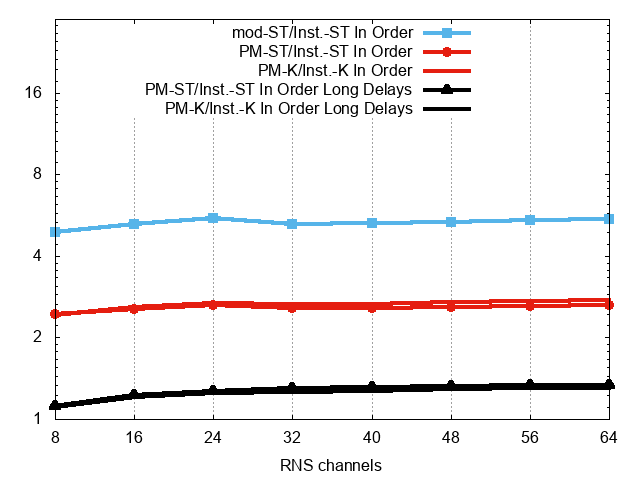}
    \caption{RNS modular multiplication Speed-Up of Inst. over the C modulo operation and pseudo-Mersenne reduction (PM), In Order processor model}
    \label{fig:SpeedUp_mod-InstIO}
\end{figure}

Figure \ref{fig:SpeedUp_mod-InstIO} summarizes the results in terms of relative speed-ups. Considering the versions with the C language modulo operation and comparing it with our instruction equipped processor implementation, the speed-ups are weakly correlated with the RNS channel number.

The use of the C modulo operation gives the slowest configurations. This operation is basically a division which is the slowest arithmetic operator implemented in the processors. The modular reduction with pseudo-Mersenne is always slower than the versions using our instructions for word-size modular operations. We notice that:
	\begin{itemize}
		\item with the delays for word-size operations \texttt{addmod}=2, \texttt{mulmod}=4, the best speed-up over pseudo-Mersenne is of 2.76. It is achieved with the Kaxamura \textsl{et al.}'s configuration, while the gain in the Szabo-Tanaka case is slightly below, see red lines.
		\item with the \textsl{long delays} case (\texttt{addmod}=4, \texttt{mulmod}=9), the speed-ups are around 1.8 at best, see black lines.
        \item the improvement provided by our instructions over pseudo-Mersenne is similar in the Szabo-Tanaka and the Kawamura \textsl{et al.} case (see red curves). We observe the same in the \textsl{long delays} simulations (black curves).
	\end{itemize}

The second part of the table \ref{tab:perf_synth} shows the speed-ups for 64-channel RNS focusing on the benefit of the versions using our instructions for word-size modular operations. Whatever the architecture version (In or Out of Order, \textsl{long delays}), the speed-ups range from 1.87 (Szabo and Tanaka, Pseudo-Mersenne versus Inst., In Order and slowest delays) up to 19 (Szabo and Tanaka, Modulo versus Inst.). The most significant speed-ups are between the fastest conventional version (Pseudo-Mersenne and Kawamura) and the fastest with the word-size modular operation Instruction (Kawamura again):
\begin{itemize}
    \item RISC-V Out of Order \texttt{mulmod} delay: 4 \texttt{addmod} delay: 2, the speed-up is 3.06;
    \item RISC-V Out of Order \texttt{mulmod} delay: 9 \texttt{addmod} delay: 4, the speed-up is 2.30.
\end{itemize}

\begin{table*}[ht]
    \centering
    \begin{tabular}{|l|rcl|c|}
\hline
    \multicolumn{5}{|c|}{\bf Base extension comparison }\\
\hline
    \hline
    \multicolumn{5}{|c|}{RISC-V In Order \hspace{5mm} \texttt{mulmod} delay: 4 \hspace{5mm} \texttt{addmod} delay: 2}\\
\hline
Mod. Operation & \multicolumn{3}{|c|}{Compared base extension methods} & Ratios \\
\hline
Pseudo-Mersenne & 	\multirow{2}*{\begin{tabular}{c} Szabo-Tanaka \end{tabular}} & \multirow{2}*{\begin{tabular}{c}  vs.  \end{tabular}} & \multirow{2}*{\begin{tabular}{l} Kawamura \textsl{et al.} 
\end{tabular}}    &  1.30 \\
Instruction &  &  &   & 1.37 \\
\hline
\hline
\hline
   \multicolumn{5}{|c|}{\bf Modulo operation comparison}\\
\hline
    \hline
Base extension & \multicolumn{3}{|c|}{Compared modulo operation methods} & Ratios \\
\hline
    \multicolumn{5}{|c|}{RISC-V In Order \hspace{5mm} \texttt{mulmod} delay: 4 \hspace{5mm} \texttt{addmod} delay: 2}\\
\hline
 Szabo-Tanaka &   Modulo & vs. &  Instruction &  5.46 \\
 Szabo-Tanaka & 	 Pseudo-Mersenne &   vs. &  Instruction &  2.63 \\

Kawamura \textsl{et al.}  &  Pseudo-Mersenne &  vs. &  Instruction  &  2.76 \\

\hline
\hline
\multicolumn{5}{|c|}{RISC-V In Order \hspace{5mm} \texttt{mulmod} delay: 9 \hspace{5mm} \texttt{addmod} delay: 4} \\
\hline
 Szabo-Tanaka &  Pseudo-Mersenne & vs. &   Instruction  &  1.87 \\

 Kawamura \textsl{et al.} &  Pseudo-Mersenne & vs. &  Instruction &  1.92 \\
\hline
\hline
    \multicolumn{5}{|c|}{RISC-V Out of Order \hspace{5mm} \texttt{mulmod} delay: 4 \hspace{5mm} \texttt{addmod} delay: 2}\\
\hline
 Szabo-Tanaka & Modulo & vs. &   Instruction &   19 \\
 Szabo-Tanaka & Modulo & vs. &   Pseudo-Mersenne &  4.53 \\

Szabo-Tanaka	 & Pseudo-Mersenne &  vs. & Instruction &  4.19 \\

 Kawamura \textsl{et al.} &  Pseudo-Mersenne &  vs. &  Instruction  &  3.06 \\
\hline
\hline
\multicolumn{5}{|c|}{RISC-V Out of Order \hspace{5mm} \texttt{mulmod} delay: 9 \hspace{5mm} \texttt{addmod} delay: 4} \\
\hline
  Szabo-Tanaka &  Pseudo-Mersenne & vs. &  Instruction  &   2.74 \\

 Kawamura \textsl{et al.} &  Pseudo-Mersenne &  vs. &  Instruction  &  2.30 \\
\hline
    \end{tabular}
    \caption{Speed ratio of 64-channel RNS modular multiplication with several word modular operations and base extensions, In Order, Out of Order RISC-V processors}
    \label{tab:perf_synth}
\end{table*}

These values show the interest of the implementation of the word-size modular operation Instructions in our simulation, and this is the motivation to continue this work in future hardware implementations.

\subsection{Architectures comparison}

\begin{figure}[ht]
    \centering
\includegraphics[width=0.95\columnwidth]{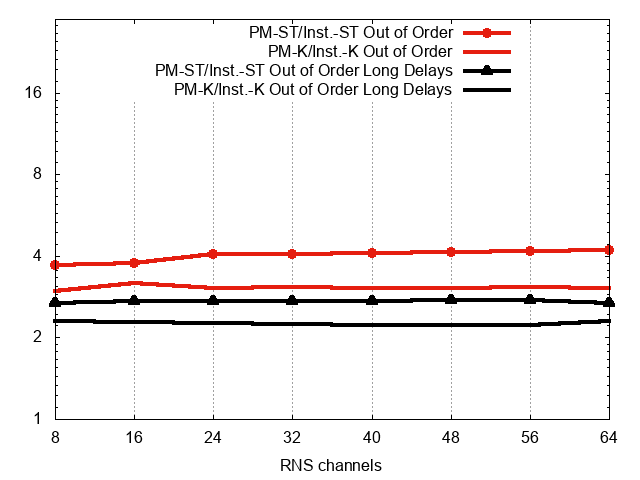}
    \caption{RNS modular multiplication Speed-Up of Inst. over the Pseudo-Mersenne reduction (PM), Out of Order processor model}
    \label{fig:SpeedUp_mod-instOOO}
\end{figure}

Figure \ref{fig:SpeedUp_mod-instOOO} shows the relative speed-ups of various combinations of modular operation and base extension methods on Out of Order processors. It highlights the improvement provided by the use  of the modular instructions \texttt{addmod} and \texttt{mulmod}. 

\subsubsection{Out of order processors}

Out of Order processors offer the opportunity to compute simultaneously in several processing units. This leads to a better use of the hardware and as a consequence, a better use of our instructions. 
For instance, modular multiplication with word-size modular operation performed with our instructions and Kawamura's base extension is 25.79 times faster than using the modulo operator and Szabo-Tanaka base extension (see Tab. \ref{tab:perf_ratios}). Compared to the fastest implementation that uses pseudo-Mersennes moduli and Kawamura's base extension, the implementation with our modular operation instructions is 3.06 times faster (Tab. \ref{tab:perf_synth}).

\subsubsection{Long delays case}

Our simulations are based on the assumption that word-size modular additions and multiplications can be computed with a delay of 2 and 4 cycles respectively. Although we believe that these delays are feasible, they could be much longer in a real implementations. Thus, we have estimated the delay of RNS modular multiplication using word-size modular additions and multiplications with a delay of up to 4 and 9, respectively.

In both cases, the speed gain through the use of our instructions remains interesting (see Tab. \ref{tab:perf_synth}).
Compared to the pseudo-Mersenne and Kawamura's case, the timing improvement for In Order processors is 92\%, and more than 2 times faster in the Out of Order case.

\subsubsection{Comparison with x86}

As a sake of comparison, we also simulated implementations of the RNS modular multiplication compiled for the Intel x86 ISA. The simulated architectures have the same cache, memory and frequency parameters. Both RISC-V and x86 architecture have two integer ALU of delay 1 and one integer multiplier of delay 3. The RISC-V has one modular adder of delay 2 and one modular multiplier of delay 4. The x86 ISA does not make use of specific word-size modular operator. The x86 binaries have been compiled with the same version of \texttt{gcc} and the same options than for the RISC-V target.
We evaluated the timings for both In Order and Out of Order GEM5 processor models.

 {\small
\begin{table}[ht]
    \centering
    \begin{tabular}{|l|rl|c|}
    \hline
   Proc. & Mod. operations & Base extension & Cycles \\
    \hline
         \multicolumn{4}{|c|}{In Order}\\
\hline
x86 & Pseudo-Mersenne & Szabo-Tanaka & 892302 \\
x86 & Pseudo-Mersenne & Kawamura \textsl{et al.} & 615359 \\
\hline
 RISC-V & Pseudo-Mersenne & Szabo-Tanaka &  446603 \\
 RISC-V & Instruction & Szabo-Tanaka & 177255 \\
 
 RISC-V  &  Pseudo-Mersenne & Kawamura \textsl{et al.} & 342415 \\
 RISC-V  & Instruction & Kawamura \textsl{et al.} & 135682 \\
\hline
     \multicolumn{4}{|c|}{Out of Order}\\
     \hline
x86 & Pseudo-Mersenne & Szabo-Tanaka & 484142 \\
x86 & Pseudo-Mersenne & Kawamura \textsl{et al.} & 361170 \\
\hline
 RISC-V & Pseudo-Mersenne & Szabo-Tanaka &  249718 \\
 RISC-V & Instruction & Szabo-Tanaka & 59548 \\
 
 RISC-V  &  Pseudo-Mersenne & Kawamura \textsl{et al.} & 134120 \\
 RISC-V  & Instruction & Kawamura \textsl{et al.} & 43865 \\

\hline
    \end{tabular}
    \caption{Number of cycles of 64-channel RNS modular multiplication with several word modular operations and base extensions, RISC-V and x86 processors}
    \label{tab:x86Cycles}
\end{table}
}

Table \ref{tab:x86Cycles} summarizes the cycles required for RNS modular multiplication in both architectures using various word-size modular operation and base extension methods. These results cover both In Order and Out of Order processors. 

In this experiment, it can be observed that the fastest RISC-V implementation of the RNS modular multiplication with our modular operation instructions requires fewer cycles than the fastest x86 implementation. In the case of In Order processors, it takes 4.5 times less cycles. In the Out of Order case, the ratio is 8 times less. 

Without using our word-size modular instructions, the best combination is to use pseudo-Mersenne moduli and Kawamura \textsl{et al.} base extension. In this case, the RISC-V RNS modular multiplication needs
79\% less cycles in In Order processors. It is 3.61 times faster for Out of Order processors.

\section{Conclusion}

In this paper, we have presented the impact on the performance of RISC-V dedicated word-size modular operation instructions for RNS modular multiplication with large modulus. We have studied the following configurations:

\begin{itemize}
    \item 2 modular multiplication variants:
        \begin{itemize}
            \item Szabo-Tanaka
            \item Kawamura
        \end{itemize}
    \item 3 word size modular operation variants:
        \begin{itemize}
            \item "C" compiled modulo
            \item Pseudo-Mersenne moduli
            \item our proposed new RISC-V Instructions, with two variants of delay
        \end{itemize}
    \item 2 RISC-V configurations
        \begin{itemize}
            \item In Order processor
            \item Out Of Order processor
        \end{itemize}
\end{itemize}

These combinations have been used to experiment with sequential RNS modular multiplications whose precision ranges from 8 to 64 64-bit RNS channels, or 512 to 4096 bits for the modulo size. The simulations were performed with the GEM5 simulator. The total number of simulations is almost 3000. We measured the performance in clock cycles for each of these configurations. We also ran the corresponding implementation on the x86-64 GEM5 simulator.

The use of specific instructions for word-size modular operations greatly improves the Residue Number Systems computation speed on both In Order and Out of Order RISC-V processors. The benefit is greater on Out of Order architectures due to the parallel nature of Residue Number Systems. Compared to the fastest implementation of RNS modular multiplication using pseudo-Mersenne moduli and the Kawamura \textsl{et al.} base extension, using our instructions yields implementations that are up to 3 times faster. The gain remains important for word-size modular operators with long delays, even with modular multiplier that has twice the delay of multipliers.

This shows that the use of specific instructions for word-size modular operations has a significant impact on the performance of the software implementation of RNS operations.

\paragraph{Future work}
This result motivates future hardware implementations of word-size modular operators for RISC-V processors.

We will also extend the simulation exploration to other configurations such as investigating different instruction strategies that use three instead of four registers. In the medium term, we also plan to implement corresponding vector instruction sets to take advantage of the natural ability to parallelization of the RNS. Our goal is to achieve, if possible, competitive performance levels for software RNS implementations on RISC-V-like platforms. This may also enable secure and randomized implementations of large modular operations for cryptographic use cases.

\bibliographystyle{plain}
\bibliography{biblio.bib}

\begin{thebibliography}{10}

\bibitem{akram2019validation}
Ayaz Akram and Lina Sawalha.
\newblock Validation of the gem5 simulator for x86 architectures.
\newblock In {\em 2019 IEEE/ACM Performance Modeling, Benchmarking and
  Simulation of High Performance Computer Systems (PMBS)}, pages 53--58. IEEE,
  2019.

\bibitem{alkim2020isa}
Erdem Alkim, H{\"u}lya Evkan, Norman Lahr, Ruben Niederhagen, and Richard
  Petri.
\newblock {ISA} extensions for finite field arithmetic accelerating kyber and
  newhope on {RISC-V}.
\newblock {\em IACR Transactions on Cryptographic Hardware and Embedded
  Systems}, 2020(3), 2020.

\bibitem{abs_12}
S.~Antão, J.-C. Bajard, and L.~Sousa.
\newblock {RNS} based elliptic curve point multiplication for massive parallel
  architectures.
\newblock {\em The Computer Journal}, 55(5):629--647, 2012.

\bibitem{aoki2022efficient}
Daichi Aoki, Kazuhiko Minematsu, Toshihiko Okamura, and Tsuyoshi Takagi.
\newblock Efficient word size modular multiplication over signed integers.
\newblock In {\em 2022 IEEE 29th Symposium on Computer Arithmetic (ARITH)},
  pages 94--101. IEEE, 2022.

\bibitem{RiscvManualVector}
Krste Asanovi.
\newblock {\em {RISC-V "V"} Vector Extension, Version 0.9-draft-1535cc0}.
\newblock EECS Department, University of California, Berkeley, 2019.

\bibitem{bajard1998rns}
J-C Bajard, L-S Didier, and Peter Kornerup.
\newblock An {RNS} montgomery modular multiplication algorithm.
\newblock {\em IEEE Transactions on Computers}, 47(7):766--776, 1998.

\bibitem{bde_13}
J.-C. Bajard, S.~Duquesne, and M.~Ercegovac.
\newblock Combining leak-resistant arithmetic for elliptic curves defined over
  $f_p$.
\newblock {\em Publications Mathématiques de Besançon. Algrèbre et Théorie
  des Nombres}, pages 67--87, 2013.
\newblock ISSN: 1958-7236.

\bibitem{behz_16}
J.-C. Bajard, Julien Eynard, Anwar Hasan, and Vincent Zucca.
\newblock A full {RNS} variant of fv like somewhat homomorphic encryption
  schemes.
\newblock In {\em SAC 2016, Selected Areas in Cryptography, St. John's,
  Newfoundland and Labrador, Canada}, 2016.

\bibitem{BajardI04}
J.-C. Bajard and L.~Imbert.
\newblock A full {RNS} implementation of {RSA}.
\newblock {\em IEEE Transactions on Computers}, 53(6):769--774, 2004.

\bibitem{binkert2011gem5}
Nathan Binkert, Bradford Beckmann, Gabriel Black, Steven~K Reinhardt, Ali
  Saidi, Arkaprava Basu, Joel Hestness, Derek~R Hower, Tushar Krishna, Somayeh
  Sardashti, et~al.
\newblock The gem5 simulator.
\newblock {\em ACM SIGARCH computer architecture news}, 39(2):1--7, 2011.

\bibitem{butko2012accuracy}
Anastasiia Butko, Rafael Garibotti, Luciano Ost, and Gilles Sassatelli.
\newblock Accuracy evaluation of gem5 simulator system.
\newblock In {\em 7th International workshop on reconfigurable and
  communication-centric systems-on-chip (ReCoSoC)}, pages 1--7. IEEE, 2012.

\bibitem{cmzt2015}
Chip-Hong Chang, Amir~Sabbagh Molahosseini, Azadeh Alsadat~Emrani Zarandi, and
  Tian~Fatt Tay.
\newblock Residue number systems: A new paradigm to datapath optimization for
  low-power and high-performance digital signal processing applications.
\newblock {\em IEEE Circuits and Systems Magazine}, 15(4):26--44, 2015.

\bibitem{cui2023risc}
Enfang Cui, Tianzheng Li, and Qian Wei.
\newblock {RISC-V} instruction set architecture extensions: A survey.
\newblock {\em IEEE Access}, 11:24696--24711, 2023.

\bibitem{didier2022software}
Laurent-St{\'e}phane Didier, Jean-Marc Robert, Fangan~Yssouf Dosso, and Nadia
  El~Mrabet.
\newblock A software comparison of {RNS} and {PMNS}.
\newblock In {\em 2022 IEEE 29th Symposium on Computer Arithmetic (ARITH)},
  pages 86--93. IEEE, 2022.

\bibitem{endo2014micro}
Fernando~A Endo, Damien Courouss{\'e}, and Henri-Pierre Charles.
\newblock Micro-architectural simulation of in-order and out-of-order arm
  microprocessors with gem5.
\newblock In {\em 2014 international conference on embedded computer systems:
  Architectures, modeling, and simulation (SAMOS XIV)}, pages 266--273. IEEE,
  2014.

\bibitem{fritzmann2020extending}
Tim Fritzmann, Georg Sigl, and Johanna Sep{\'u}lveda.
\newblock Extending the {RISC-V} instruction set for hardware acceleration of
  the post-quantum scheme {LAC}.
\newblock In {\em 2020 Design, Automation \& Test in Europe Conference \&
  Exhibition (DATE)}, pages 1420--1425. IEEE, 2020.

\bibitem{fritzmann2020risq}
Tim Fritzmann, Georg Sigl, and Johanna Sep{\'u}lveda.
\newblock {RISQ-V}: Tightly coupled {RISC-V} accelerators for post-quantum
  cryptography.
\newblock {\em IACR Transactions on Cryptographic Hardware and Embedded
  Systems}, pages 239--280, 2020.

\bibitem{gar_56}
H.~L. Garner.
\newblock The residue number system.
\newblock {\em IRE Transactions on Electronic Computers}, EL 8(6):140–147,
  1959.

\bibitem{gnu_mp}
Torbjörn Granlund and al.
\newblock {GNU} multiple precision arithmetic library 6.1.2.
\newblock \url{https://gmplib.org/}.

\bibitem{huang2022improved}
Junhao Huang, Jipeng Zhang, Haosong Zhao, Zhe Liu, Ray~CC Cheung,
  {\c{C}}etin~Kaya Ko{\c{c}}, and Donglong Chen.
\newblock Improved plantard arithmetic for lattice-based cryptography.
\newblock {\em IACR Transactions on Cryptographic Hardware and Embedded
  Systems}, 2022(4):614--636, 2022.

\bibitem{huang2023yet}
Junhao Huang, Haosong Zhao, Jipeng Zhang, Wangchen Dai, Lu~Zhou, Ray~CC Cheung,
  Cetin~Kaya Koc, and Donglong Chen.
\newblock Yet another improvement of plantard arithmetic for faster kyber on
  low-end 32-bit iot devices.
\newblock {\em arXiv preprint arXiv:2309.00440}, 2023.

\bibitem{karabulut2020rantt}
Emre Karabulut and Aydin Aysu.
\newblock {RANTT}: {A RISC-V} architecture extension for the number theoretic
  transform.
\newblock In {\em 2020 30th International Conference on Field-Programmable
  Logic and Applications (FPL)}, pages 26--32. IEEE, 2020.

\bibitem{Cox-Rower00}
Shinichi Kawamura, Masanobu Koike, Fumihiko Sano, and Atsushi Shimbo.
\newblock Cox-rower architecture for fast parallel montgomery multiplication.
\newblock In Bart Preneel, editor, {\em Advances in Cryptology --- EUROCRYPT
  2000}, pages 523--538, Berlin, Heidelberg, 2000. Springer Berlin Heidelberg.

\bibitem{kiaei2023architecture}
Pantea Kiaei, Thomas Conroy, and Patrick Schaumont.
\newblock Architecture support for bitslicing.
\newblock {\em IEEE Transactions on Emerging Topics in Computing},
  11(2):497--510, 2023.

\bibitem{knuth2014art}
Donald~E Knuth.
\newblock {\em Art of computer programming, volume 2: Seminumerical
  algorithms}.
\newblock Addison-Wesley Professional, 2014.

\bibitem{lesavourey2017efficient}
Andrea Lesavourey, Christophe Negre, and Thomas Plantard.
\newblock Efficient leak resistant modular exponentiation in rns.
\newblock In {\em 2017 IEEE 24th Symposium on Computer Arithmetic (ARITH)},
  pages 156--163. IEEE, 2017.

\bibitem{mallasen2022percival}
David Mallas{\'e}n, Raul Murillo, Alberto~A Del~Barrio, Guillermo Botella, Luis
  Pi{\~n}uel, and Manuel Prieto-Matias.
\newblock {PERCIVAL}: open-source posit {RISC-V} core with quire capability.
\newblock {\em IEEE Transactions on Emerging Topics in Computing},
  10(3):1241--1252, 2022.

\bibitem{menezes2018handbook}
Alfred~J Menezes, Paul~C Van~Oorschot, and Scott~A Vanstone.
\newblock {\em Handbook of applied cryptography}.
\newblock CRC press, 2018.

\bibitem{mohan2017rns}
PV~Ananda Mohan, PK~Meher, and T~Stouraitis.
\newblock {\em Arithmetic circuits for DSP applications}, chapter RNS-Based
  arithmetic circuits and applications, pages 186--236.
\newblock John Wiley \& Sons, 2017.

\bibitem{montMult_85}
Peter~L. Montgomery.
\newblock Modular multiplication without trial division.
\newblock {\em Mathematics of Computation}, 44(170):519--521, 1985.

\bibitem{Plantard21}
Thomas Plantard.
\newblock Efficient word size modular arithmetic.
\newblock {\em IEEE Transactions on Emerging Topics in Computing},
  9(3):1506--1518, 2021.

\bibitem{posch1995modulo}
Karl~C Posch and Reinhard Posch.
\newblock Modulo reduction in residue number systems.
\newblock {\em IEEE Transactions on Parallel and Distributed Systems},
  6(5):449--454, 1995.

\bibitem{qiu2023gem5tune}
Yudi Qiu, Tao Huang, Yuxin Tang, Yanwei Liu, Yang Kong, Xulin Yu, Xiaoyang
  Zeng, and Yibo Fan.
\newblock Gem5tune: A parameter auto-tuning framework for gem5 simulator to
  reduce errors.
\newblock {\em IEEE Transactions on Computers}, 2023.

\bibitem{rawat2017vector}
Hemendra Rawat and Patrick Schaumont.
\newblock Vector instruction set extensions for efficient computation of
  keccak.
\newblock {\em IEEE Transactions on Computers}, 66(10):1778--1789, 2017.

\bibitem{riscvToolChain}
riscv collab.
\newblock {RISC-V GNU} compiler toolchain.
\newblock \url{https://github.com/riscv-collab/riscv-gnu-toolchain}, 2022.

\bibitem{roelke2017risc5}
Alec Roelke and Mircea~R Stan.
\newblock Risc5: Implementing the {RISC-V ISA} in gem5.
\newblock In {\em First Workshop on Computer Architecture Research with RISC-V
  (CARRV)}, volume~7, 2017.

\bibitem{shenoy-kumaresan89}
A.P. Shenoy and R.~Kumaresan.
\newblock Fast base extension using a redundant modulus in {RNS}.
\newblock {\em IEEE Transactions on Computers}, 38(2):292--297, 1989.

\bibitem{szabo-tanaka67}
Nicholas~S Szabo and Richard~I Tanaka.
\newblock {\em Residue arithmetic and its applications to computer technology}.
\newblock New York: McGraw-Hill, 1967.

\bibitem{tay2017fault}
Thian~Fatt Tay and Chip-Hong Chang.
\newblock {\em Embedded systems design with special arithmetic and number
  systems}, chapter Fault-tolerant computing in redundant residue number
  system, pages 65--88.
\newblock Springer, 2017.

\bibitem{taylor84}
Taylor.
\newblock Residue arithmetic a tutorial with examples.
\newblock {\em Computer}, 17(5):50--62, 1984.

\bibitem{tiwari2021peri}
Sugandha Tiwari, Neel Gala, Chester Rebeiro, and V~Kamakoti.
\newblock {PERI}: A configurable posit enabled risc-v core.
\newblock {\em ACM Transactions on Architecture and Code Optimization (TACO)},
  18(3):1--26, 2021.

\bibitem{valueva2020application}
Maria~V Valueva, NN~Nagornov, Pavel~Alekseevich Lyakhov, Georgii~V Valuev, and
  Nikolay~I Chervyakov.
\newblock Application of the residue number system to reduce hardware costs of
  the convolutional neural network implementation.
\newblock {\em Mathematics and computers in simulation}, 177:232--243, 2020.

\bibitem{walker2018hardware}
Matthew Walker, Sascha Bischoff, Stephan Diestelhorst, Geoff Merrett, and
  Bashir Al-Hashimi.
\newblock Hardware-validated {CPU} performance and energy modelling.
\newblock In {\em 2018 IEEE International Symposium on Performance Analysis of
  Systems and Software (ISPASS)}, pages 44--53. IEEE, 2018.

\bibitem{RiscvManual1}
Andrew Waterman and Krste Asanovi.
\newblock {\em The {RISC-V} Instruction Set Manual Volume I: Unprivileged {ISA}
  version 20191213}.
\newblock EECS Department, University of California, Berkeley, 2019.

\end{thebibliography}

\end{document}